\documentclass[journal=nalefd,manuscript=letter,super=false]{achemso}
\setcitestyle{numbers,round}
\setkeys{acs}{keywords = true} 
\setkeys{acs}{maxauthors=0}
\setkeys{acs}{doi = true}
\setkeys{acs}{hyperref = true}
\usepackage{chemformula} 
\usepackage[T1]{fontenc} 
\usepackage{hyperref}
\usepackage{lineno}

\newcommand*{\doi}[1]{\href{http://doi.org/#1}{#1}}

\author{Franco N. Di Rino}
\email{dirino@fzu.cz}

\author{Tim Verhagen}
\email{verhagen@fzu.cz}
\affiliation{Institute of Physics, Czech Academy of Sciences, Na Slovance 2, Prague 8, 182 00, Czech Republic}

\title[Switchable Topological Polar Textures in Freestanding Ultrathin Ferroelectric Oxides]{Switchable Topological Polar Textures in Freestanding Ultrathin Ferroelectric Oxides}

\begin{document}


\begin{abstract}
The rapid expansion of two-dimensional materials has recently extended to freestanding complex oxides, opening new opportunities for nanoscale ferroic design. Using first-principles-based atomistic simulations, we demonstrate that ultrathin freestanding ferroelectric layers host a diverse landscape of polar states. Above a critical thickness, electrostatic confinement stabilizes a vortex–labyrinthine regime with liquid-like out-of-plane domains and long-range orientational order, which upon cooling evolves into two nearly degenerate topological configurations: a wave–helix texture and a chiral bubbles phase. Remarkably, these states are deterministically and reversibly interconverted by static and THz electric fields, enabling ultrafast electrical control of topological states. The small energy separation between phases creates a programmable energy landscape, establishing freestanding ferroelectric nanolayers as reconfigurable platforms for topological nanoelectronics without structural twisting or interface engineering.
\end{abstract}

\section{}
Ferroelectricity in low-dimensional systems favors the formation of a wide variety of topological structures, which emerge as an electrostatic mechanism to minimize the depolarizing energy associated with bound charges at the surfaces of uniformly polarized nanostructures \cite{Das2019,Das_2020,Shao2023,Junquera_2023,Lukyanchuk_2025}. These textures can be designed, stabilized, and reconfigured through confinement and boundary conditions, making their controlled manipulation appealing for future devices \cite{Catalan_2012,Chen_2020}. Understanding how polarization adapts under spatial confinement is therefore essential to engineer topological states and advance nanoscale ferroelectric functionality.

Traditionally, the properties of ferroelectric thin layers have been tuned through substrate engineering, where epitaxial strain, mechanical clamping, and interfacial electrostatics dictate domain formation \cite{Hu_2024,Rabe_2005,Tinte_2001,Pertsev_1988,Pertsev_1999}. While these strategies have enabled remarkable progress, they constrain ferroic behavior to substrate-bound systems, where strong mechanical and electrical boundary conditions limit the material’s response.

Recent progress in two-dimensional (2D) materials has introduced powerful experimental routes to engineer functional properties via layer stacking. Interlayer degrees of freedom such as sliding, twisting, and reconstruction can break inversion symmetry and induce ferroelectric-like behavior \cite{Wu_2021,Weston_2022,Hassan_2024,Li_2024}. Importantly, studies in the single-layer limit have revealed exceptionally rich physics spanning structural, electronic, and topological phenomena, as exemplified by graphene and other atomically thin materials \cite{Novoselov2004,Castro_2009}. These discoveries highlight how emergent behavior can arise even in the simplest, unconstrained geometries.

Inspired by these advances, similar design principles are now being extended to complex oxides, enabling the fabrication of freestanding and twisted oxide layers with atomic-level precision \cite{Fernandez_2022,Chiabrera_2022}. For instance, Sánchez-Santolino et al. reported polarization vortex-antivortex arrays in twisted freestanding BaTiO$_3$ (BTO) layers \cite{Sanchez_2024}, while related works have shown that such topological polarization textures can be tailored through twist and strain \cite{Sha_2024,Zhang_2024,Lee_2024}. Although these developments bring oxide systems closer to the concepts explored in 2D materials, research has so far focused mainly on complex stacked architectures, whereas the behavior of freestanding ultrathin oxides approaching the single-layer regime remains largely unexplored.

Freestanding ferroelectric layers thus provide an opportunity to examine polarization behavior in the absence of substrate-induced constraints. By removing external boundary effects, they provide a clean framework to explore the formation and stability of complex polarization textures in reduced dimensions. In this context, atomistic models parameterized from first-principles calculations offer a powerful tool to explore size effects, structural instabilities, dynamical behavior, and emergent polarization textures in low-dimensional, unconstrained environments. In this work, we employ a core-shell model to study the polarization patterns in freestanding BTO thin layers, revealing a variety of stable polar configurations. 

\begin{figure}
\centering
\includegraphics[width=\linewidth]{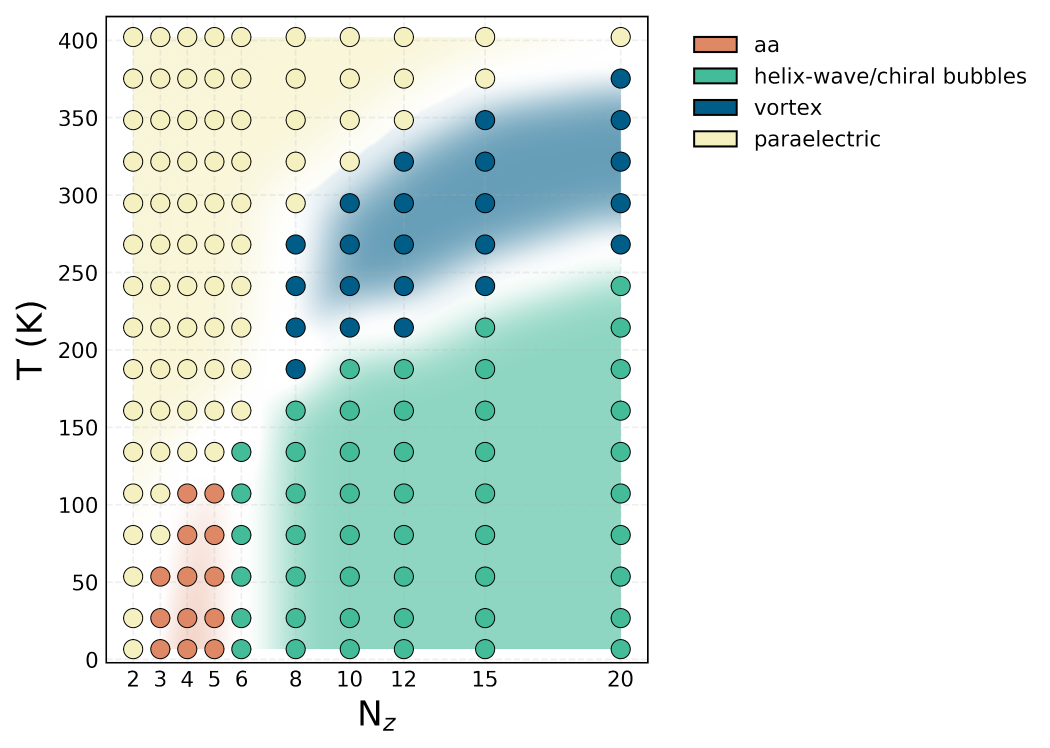}
\caption{\label{fig:phasedia} Ferroelectric phases diagram (layer thickness N$_z$ - temperature) in free standing BTO thin layers. Colored symbols represent discrete simulation results, with colors used to distinguish different phases.}
\end{figure}

To elucidate how these patterns emerge and evolve, we systematically map the polarization configurations as a function of temperature and layer thickness $N_z$, defined as the number of Ti atoms along the pseudocubic $z$ direction (Figure~\ref{fig:phasedia}). The different phases are identified based on the temperature dependence of the polarization components and lattice parameters. These distinctions provide a consistent framework to classify the different polar textures beyond visual inspection.  Starting from the paraelectric phase, we track the evolution of freestanding BTO layers of varying thicknesses upon cooling.

For ultrathin layers ($N_z \le 2$), no stable ferroelectric order develops, and the system remains dominated by thermal fluctuations, with net polarization values fluctuating around zero. 

For intermediate thicknesses ($3 \le N_z < 6$), temperature reduction stabilizes a single-domain $aa$-type ferroelectric phase, characterized by an in-plane polarization along the $\langle110\rangle$ direction. In this phase, two in-plane components of the polarization are finite, while the out-of-plane component remains suppressed.

For thicker layers ($N_z \ge 6$), the system instead organizes into three distinct nonuniform polar textures characterized by extended vortex textures.

\begin{figure}
\centering
\includegraphics[width=\linewidth]{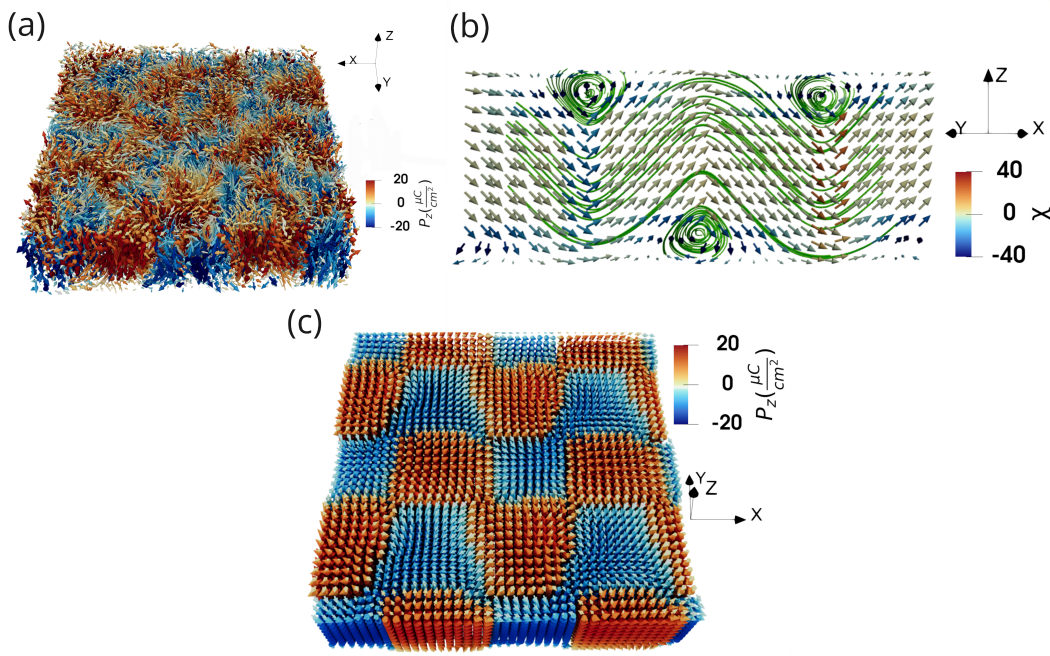}
\caption{\label{fig:states} Representative three-dimensional polarization textures stabilized at different temperatures in a freestanding BTO layer ($N_z=12$).
(a) Vortex labyrinthine structure vector field. Streamlines serve as a guide to the eye (240 K).
(b) Wave-helix (7 K). Cross-sectional view along a $\langle110\rangle$ plane. Streamlines serve as a guide to eye.
(c) Chiral bubbles (7 K) vector field exhibiting square-like domain organization with alternating $P_z$ orientation.}
\end{figure}

Just below the paraelectric transition the system forms a \textit{vortex–labyrinthine phase} (Figure~\ref{fig:states}a), where alternating out-of-plane polarized domains arrange into an irregular, liquid-like labyrinth. This phase is dominated by a single out-of-plane polarization component ($P_z$) and exhibits a pronounced local tetragonal distortion with predominantly 180° domain walls. Within this texture, the polarization rotates continuously across the mobile domain walls and defines vortex lines that fluctuate near the layer center and occasionally bend into \textit{Néel-type bubbles}. As a result, the labyrinth hosts dynamically reconfigurable vortices with mixed rotational handedness, similar to those reported in PbTiO$_3$/SrTiO$_3$ (PTO/STO) superlattices and strained ferroelectric thin layers \cite{Zubko_2016,Ortiz_2024,Nahas_2020,Nahas_2020_2,Boron_2025}.

Upon further cooling, thermal fluctuations diminish and the labyrinthine texture progressively freezes. The vortex cores shift toward the layer surfaces, giving rise to two nearly degenerate low-temperature polar textures that are largely static and characterized by finite values of all three components of the local polarization.

The \textit{wave-helix} state consists of elongated helical segments with a well-defined axis, producing stripe-like out-of-plane domains with a characteristic periodicity (Figure~\ref{fig:states}b). In contrast, the \textit{chiral bubbles} state develops square-like domains formed by bent helical cores that tend to close into toroidal loops (details of the topological analysis can be found in Section S1 in the Supporting Information), yielding bubble-like textures with a well-defined chirality  (Figure~\ref{fig:states}c).

Taken together, these temperature-driven transformations outline the phase diagram shown in Figure~\ref{fig:phasedia}, which closely resembles that obtained from thermodynamic free-energy calculations using the soft-domain analytical framework once surface-tension effects are included \cite{Kondovych_2025}. In that formulation, the polarization is represented through a small set of Fourier modes capturing the leading modulations, resulting in continuous textures where atomically sharp domain walls are not explicitly resolved. 

Consistently, both the analytical framework and our atomistic simulations indicate that surface tension plays a central stabilizing role, strongly influencing the formation and persistence of modulated polar states. In thin layers, it introduces an effective compressive stress that competes with electrostatic and elastic energies. As the thickness increases, its relative contribution evolves, modifying the balance between competing interactions and driving the crossover at $N_z=6$. Within this regime, the wave-helix and chiral bubble states accommodate similar polarization patterns with comparable energetic cost, leading to their near degeneracy.

Despite this agreement, textures such as the chiral bubbles state lie beyond the scope of the minimal soft-domain description, as their stabilization requires additional energetic contributions and the superposition of multiple wavevectors.

Having established the phase diagram and identified the relevant polarization textures, we now provide a physical interpretation of the domain morphology and energetic balance of the two competing low-temperature states.

In the wave–helix state (Figure.~\ref{fig:states}b), the in-plane polarization develops a dominant orientation along an equivalent $\langle 110 \rangle$ direction, while the out-of-plane component modulates along the layer plane.  This interplay produces stripe-like domains  with a clear domain periodicity. The associated domain walls correspond to polarization rotations close to 71\textdegree, which are mechanically compatible with rhombohedral BTO \cite{Marton2010}. While this stripe-domain configuration minimizes the number of walls and preserves the full three-component polarization, it induces extended gradients of tetragonal distortion across the layer thickness.

By contrast, the chiral bubbles state (Figure~\ref{fig:states}c) adopts a markedly different domain architecture. Here, the polarization forms square-like out-of-plane domains separated by sharp domain walls that cut across both pseudocubic in-plane directions. In three dimensions (3D), the polarization organizes into toroidal loops, where the helical cores periodically migrate between the two layer surfaces. This geometry preserves the helical character of the wave–helix state; however, instead of maintaining a well-defined axis, the cores bend and progressively tend to close into loops (further details can be seen in Figure S1 in the Supporting Information). These patterns involve local polarization rotations corresponding to 109° ferroelectric domain walls, which are also compatible with rhombohedral symmetry \cite{Marton2010}. Although a larger number of domain walls are introduced, they locally modulate the tetragonal strain and thereby relieve elastic stress without producing long-range gradients. Zero-temperature energy minimizations for a representative $N_z = 12$ system identify the wave-helix as the ground state, although the energy difference with respect to the chiral bubbles configuration is marginal ($\approx 0.3$~meV per formula unit). This near degeneracy reflects a delicate competition between distinct polarization textures and suggests that both configurations may be experimentally accessible depending on growth conditions or thermal history. 

\begin{figure}
\centering
\includegraphics[width=\linewidth]{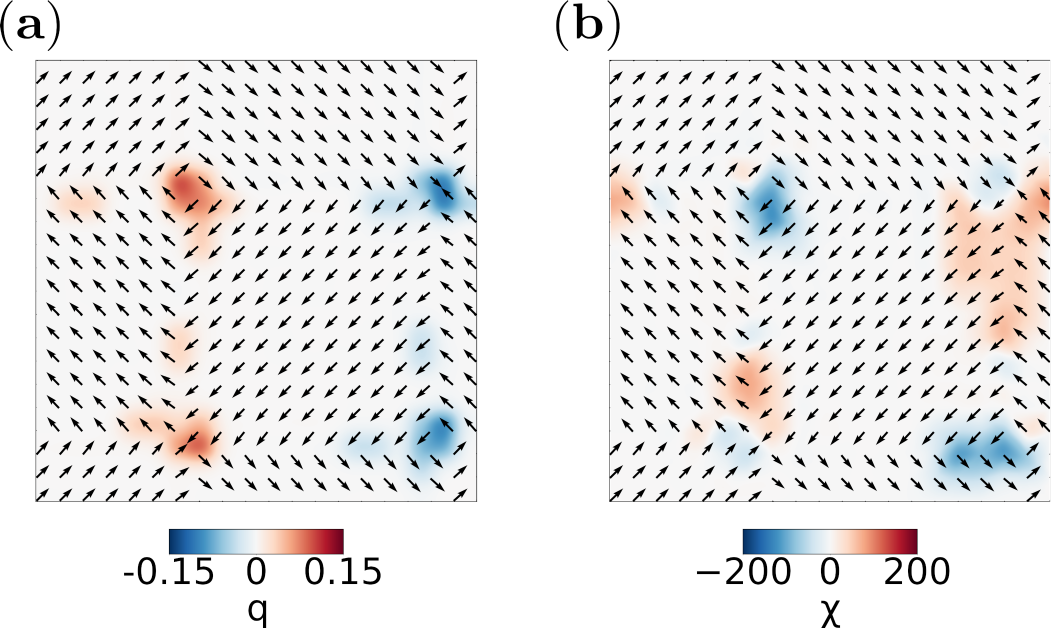}
\caption{\label{q} Cross-sectional view of the vector field in the chiral bubble phase along a $\langle001\rangle$ plane.(a) Topological density charge $\mathrm{q}$ and (b) chirality density $\chi$ map.}
\end{figure}

Among the stable low-temperature configurations, the chiral bubbles state stands out due to its intricate internal organization. To better elucidate its spatial arrangement, we consider suitable 2D projections of the polarization field. When projected onto the layer mid-plane, this configuration appears as an ordered array of alternating vortices and antivortices in the in-plane polarization components (Figure~\ref{q}). A similar vortex–antivortex ordering has been reported in twisted ferroelectric layers \cite{Sanchez_2024}, suggesting a common phenomenology emerging from modulated polarization fields. The corresponding topological charge density map, defined as $q(x, y) = (1/4\pi) \bf{n}\cdot(\partial_x \bf{n} \times   \partial_y \bf{n})$, where $\bf{n}$ is the normalized local polarization vector \cite{Nahas_2015,Gonsalves_2024}, reveals a meron-antimeron arrangement resembling those reported in bulk configurations (Figure~\ref{q}a) \cite{Bastogne2024}. Complementarily, the chirality density map, given as $\chi=\bf{P}\cdot(\nabla\times\bf{P})$ \cite{Lukyanchuk_2025}, reveals the topological handedness inherent to the individual helical cores, as visible in Figure~\ref{q}b. As discussed by Luk’yanchuk \textit{et al.} \cite{Lukyanchuk_2025} for polar skyrmions in PTO/STO heterostructures \cite{Das2019}, the Pontryagin index and the associated topological charge density characterize the topology within a given plane, but do not fully capture the intrinsically 3D  nature of chiral polarization textures. In our case, the computed topological charge represents a 2D projection of a fully 3D polarization field. The magnitude and orientation of $\mathbf{P}$ vary across the layer thickness, as its modulus is not fixed. This contrasts with magnetic systems, where the spin magnitude is typically constant. Together, the topological charge and chirality maps provide a complementary description of the chiral bubbles texture, capturing both its in-plane vortex–antivortex organization and the handedness of the underlying 3D polarization field.

While the preceding analysis highlights the mesoscopic topology of the polarization textures, it is also instructive to examine how these polarization patterns relate to the underlying local structural character of the layer. To this end, the local polarization was averaged over the simulation time and classified using a threshold-based criterion. A polarization component is considered significant when its magnitude exceeds 7~$\mu$C/cm$^2$. Accordingly, unit cells are identified as rhombohedral when all three components exceed this threshold, orthorhombic when exactly two components exceed it, tetragonal when only one component exceeds it, and paraelectric when all components remain below 7~$\mu$C/cm$^2$. The same threshold values were used for all layer thicknesses and were chosen conservatively so that all structural variants could be consistently identified. 

This analysis reveals that freestanding BTO layers undergo a sequence of structural transformations that closely mirrors the bulk rhombohedral–orthorhombic–tetragonal–paraelectric phase transitions. At low temperatures, most cells exhibit three nonzero polarization components, consistent with a rhombohedral-like local character. Upon heating, one component progressively diminishes, giving rise to an orthorhombic-like regime within a narrow temperature window. With further temperature increase, the system evolves into a predominantly tetragonal-like configuration, and eventually into a weakly polarized, fluctuation-dominated paraelectric state. Additional data supporting these phase assignments are provided in Section S2 and S3 in the Supporting Information.

These results demonstrate that BTO, even under spatial confinement, preserves a sequence of structural transformations closely resembling those of the bulk. Confinement and finite-size effects modify the detailed polarization textures, yet the overall evolution of the average structure still follows the characteristic rhombohedral-orthorhombic-tetragonal-paraelectric progression. When initialized from wave-helix configurations, the simulations do not display a clearly developed orthorhombic phase. This could simply stem from the orthorhombic regime being restricted to a very narrow temperature interval in this model.

To connect these real-space polarization textures with their characteristic length scales, we next analyze the structure factor of the out-of-plane polarization component obtained from molecular dynamics simulations for a representative $N_z = 12$ system.

\begin{figure*}
\centering
\includegraphics[width=\linewidth]{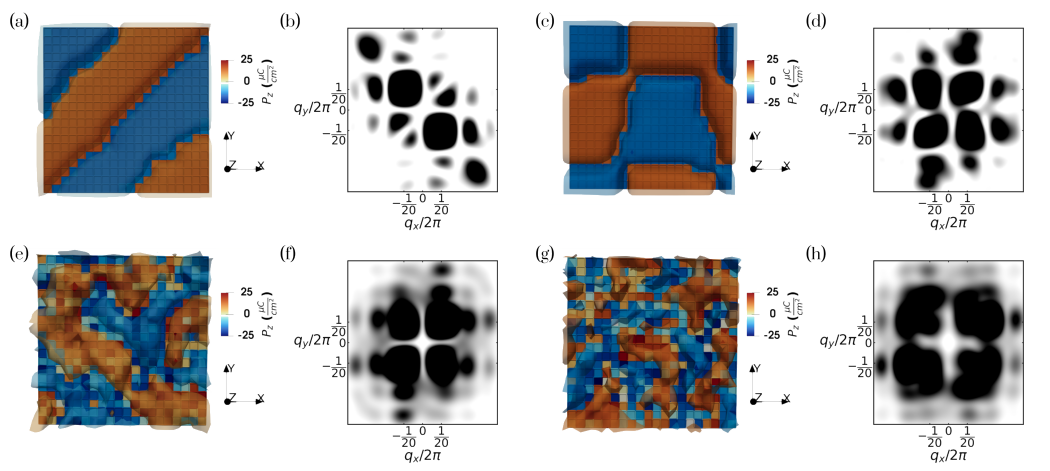}
\caption{ Top-view snapshots of the out of plane polarization component ($P_z$) of the layer. Panels (a), (c), (e), and (g) show instantaneous configurations at 7 K (stripe domain), 7 K (chiral bubbles domain), 300 K, and 325 K, respectively. The color lobes correspond to contour regions with similar $P_z$ values. Panels (b), (d), (f), and (h) display the corresponding time averaged structure factors of the out-of-plane polarization component $S(P_z)$}
\label{fig:sf}
\end{figure*}

 This statistical measure, previously applied to PTO/STO superlattices \cite{Ortiz_2024}, provides a sensitive probe of spatial correlations and domain morphology. Representative snapshots and schematic illustrations are shown in Figure~\ref{fig:sf}.

In the low-temperature regime, two distinct stable configurations can be identified. The first corresponds to a pinned stripe-like domain pattern (Fig \ref{fig:sf}a, b), whose reciprocal-space signature consists of two sharp diagonal lobes in the structure factor at $\boldsymbol{q} = \left( \mp\frac{1}{d}, \pm\frac{1}{d} \right)$ u.c.$^{-1}$ (with $d = 20$ being the domain periodicity), together with additional low-order harmonics. These secondary maxima originate from the diagonal domain orientation, which precludes a simple two-domain configuration and accommodates multiple domains within the simulation cell. This anisotropic scattering reflects a static twofold-symmetric arrangement of alternating out-of-plane polarizations oriented along the $\langle 1\overline{1}0 \rangle$ / $\langle \overline{1}10 \rangle$ directions, although an equivalent variant with lobes along the opposite diagonals ($\langle 110 \rangle$ / $\langle\overline{11}0 \rangle$) may also occur depending on the initial conditions or selected domain variant.

The second configuration, the chiral bubbles state (Figure~\ref{fig:sf}c, d), displays a structure factor with four well-defined rectangular lobes distributed along the diagonals at $q_x = \pm 1/20$ and $q_y = \pm 1/20$. This pattern originates exclusively from the modulation of the out-of-plane polarization component $P_z$. The peak positions encode the periodicity and preferred orientation of the $P_z$ variations, while the rectangular shape of the lobes reflects the in-plane anisotropy of the domains, which are slightly elongated along one direction. We find that this domain anisotropy is robust with respect to the lateral simulation cell size, as it is consistently observed across different system sizes at fixed thickness. It likely originates from the fully three-dimensional character of the polarization field in the chiral bubble phase, which constrains the domain morphology and disfavors highly symmetric in-plane arrangements. At the same time, finite-size and commensurability effects associated with the simulation cell may also influence the precise domain shape.

Upon heating, both configurations evolve toward a common high-temperature regime (Figure~\ref{fig:sf}e, f), where the structure factor becomes more diffuse while retaining an approximate fourfold symmetry along equivalent $\langle110\rangle$ directions.
This behavior reflects the loss of long-range translational order while preserving orientational correlations, consistent with a liquid-like regime exhibiting tetratic orientational symmetry.
Such fourfold orientational order, distinct from the sixfold symmetry characteristic of hexatic phases, reflects the influence of the underlying perovskite lattice. Importantly, in contrast to the squaric ordering reported in engineered ferroelectric heterostructures \cite{Ortiz_2024}, where the dominant correlations are aligned with the principal crystallographic axes, here the maxima of the structure factor lie along the diagonal $\langle110\rangle$ directions, indicating a rotation of the preferred modulation by 45°. 

In real space, this regime is accompanied by enhanced domain dynamics, as polarization domains begin to meander, merge, and undergo coherent flips, giving rise to a fluctuating labyrinthine state.
The emergence of this behavior in a single-component BTO layer evidences an intrinsic form of frustrated ferroelectric order.
Although labyrinthine patterns lack translational order by construction, the observed phase demonstrates that these structures nonetheless retain a preferred orientational tendency.
Comparable fourfold diffuse scattering has been reported in two-dimensional antiferromagnetic systems \cite{Abutbul_2022}, where competing interactions give rise to a state with no translational order but a well-defined orientational anisotropy.

At higher temperatures (Figure~\ref{fig:sf}g, h), the four lobes broaden and lose contrast, reflecting the progressive loss of spatial correlations and the emergence of an isotropic liquid-like behavior. In this regime, polarization fluctuations dominate, and the system evolves into a fully disordered state, marking the eventual loss of ferroic order.

It is worth noting that the orientational dynamics are effectively confined to the layer plane. The polarization remains coherently aligned across the layer thickness, with no mixing of directions along the out-of-plane axis due to strong electrostatic and elastic couplings. As a result, the system can be described as a two-dimensional field in which the relevant degree of freedom is the in-plane orientation of the domains.

Finally, we explore the possibility of switching between the two low-temperature topological states: the chiral bubbles state and the wave–helix state. Starting from the chiral bubbles configuration, a static or low-frequency electric field applied along any equivalent  $\langle110\rangle$ directions, sufficiently strong to break the topological protection ($\approx0.02 V/ $\AA$ $ depending on thickness), aligns the polarization along a $\langle110\rangle$ direction. This process drives the system into a wave–helix pattern, which remains robust upon field removal.

Achieving the reverse process, from the wave–helix back to the chiral bubbles state is more intricate. We demonstrate that time-dependent electric fields applied along $\langle110\rangle$, utilizing THz-frequency Gaussian pulses, can trigger a reorganization of the polarization field. This stimulus effectively transforms the stripe-like wave–helix morphology back into the chiral bubbles configuration.

For instance, in a $N_z=12$ system at 55 K, a Gaussian pulse
$E_0 e^{-( (t-t_0)/ \tau)^2} \sin(\omega t)$ with frequency $\omega=5$ THz, amplitude $E_0=0.01$ V/\AA, temporal center $t_0=100$ ps, and pulse duration $\tau=40$ ps led to the formation of stable bubble domains after 240 ps of simulation. Comparable responses were obtained across different temperatures and initial states, indicating that the effect is robust. $\omega$ and $E_0$ influence the bubble size and in-plane rotation, while the pulse duration primarily determines the time required to pump the vibrational modes involved in the rotation dynamics. The dynamics of the field-induced switching between the chiral bubble and wave-helix phase for the $N_z=12$ system is shown in Movie S1 in the Supporting Information.

This response may arise from anharmonic coupling between optical and acoustic modes. In bulk BTO, excitation of acoustic phonons can generate strain gradients that stabilize vortex-antivortex structures \cite{Bastogne2024}. In our confined layers, although the THz excitation primarily addresses optical phonons, the system behaves effectively as bulk-like within the plane, where near-degeneracy between optical and acoustic branches enhances the effect of symmetry-allowed coupling. Such anharmonic interactions could transiently reshape the local energy landscape, facilitating polarization rotations and the reorganization of toroidal structures.

Interestingly, this field-induced reorganization is not restricted to thicknesses where bubble states appear in equilibrium. Even ultrathin layers ($3 \le N_z < 6 $ unit cells) in the in-plane $aa$ phase can develop robust vortex-antivortex-like domains under analogous time-dependent fields.  In contrast to the bubble states in thicker layers, which possess a significant out-of-plane polarization component, these flux-closure textures are characterized by a polarization vector essentially confined to the layer plane (a detailed 3D comparison is provided in Section S4 in the Supporting Information).

While the microscopic mechanism requires further clarification, these results point to nonlinear phononics as a potential pathway for dynamically engineering topological polarization states in confined ferroelectrics. Time-dependent electric fields therefore offer a viable route for manipulating, and potentially designing, topological states at the nanoscale.

Overall, our simulations show that freestanding BTO layers can sustain a wide range of topological polarization textures stabilized by electrostatic and elastic confinement. At low temperatures, two nearly degenerate states appear: a wave-helix phase and a chiral bubbles domain produced by helical cores that tend to close into loops. In 2D projections, these bubbles manifest as alternating vortex-antivortex pairs, consistent with recent observations in twisted freestanding BTO layers. As the system is heated, the in-plane polarization weakens and a vortex labyrinthine state emerges. Thermal fluctuations then drive a regime with tetratic symmetry, where positional order fades while orientational correlations persist over a finite temperature window. Although this regime shares a fourfold (squaric) symmetry with patterns reported in PTO/STO superlattices, the underlying ordering differs in that the dominant correlations are oriented along diagonal directions rather than along the principal axes, reflecting a distinct anisotropy of the polarization modulation. Finally, switching under time-dependent electric fields confirms that these low-temperature textures are not only stable but also responsive and controllable.

While our simulations capture several robust equilibrium configurations, other stable or metastable phases may still arise under conditions not examined here. In BTO, the extremely small energy differences between competing polar states make the system prone to hosting multiple nearly degenerate configurations, a hallmark of relaxor-like behavior. Understanding how such states appear and evolve under different external stimuli remains an important direction for future work.

A complementary analysis of the time-averaged local polarization shows that the system preserves the characteristic bulk-like rhombohedral-orthorhombic-tetragonal-paraelectric sequence under confinement, providing a structural counterpart to the topological transformations described above.

To conclude, these results highlight that even simple, freestanding ferroelectric layers exhibit a level of topological complexity and physical richness comparable to engineered heterostructures. Remarkably, such complexity arises without the lattice reconstruction or multilayer design required in moiré systems, underscoring their potential as minimal yet powerful, easily switchable platforms for emergent ferroelectric polarization and topological phenomena.

\section{Methods}

Molecular dynamics simulations were performed to investigate the polarization patterns and structural characteristics of freestanding BTO thin layers, using an interatomic potential derived from first-principles calculations \cite{Sepliarsky_2005,Ghosez_2022}. This framework has been extensively validated and has shown good agreement with experimental data, accurately reproducing the bulk properties of pure BTO, solid solutions, and mixed compounds \cite{Machado_2019,Sepliarsky_2023, Sepliarsky_2025}. The same theoretical scheme has also been applied to various low-dimensional BTO and PTO systems, including epitaxially strained thin layers and freestanding layers addressing surface energy effects \cite{Tinte_2001,Kondovych_2025,Stachiotti_2011,DiRino_2025}, which demonstrates the robustness and transferability of this approach.
In this model, the relative displacement between the core and shell represents the ion’s electronic polarization. Interatomic interactions include harmonic and fourth-order core-shell couplings ($k_2$ and $k_4$), long-range Coulomb forces, and short-range repulsive terms.
Short-range interactions are modeled using two types of potentials: a
Born-Mayer potential, V(r)=$Ae^{-\frac{r}{\rho}}$, for Ba-O and Ti-O, and a Buckingham
potential, V(r)=$Ae^{-\frac{r}{\rho}}+\frac{r}{C^6}$, for O-O interactions, where $r$ is the interatomic distance and 
$A$, $\rho$, and $C$ are model parameters.

The local polarization $\boldsymbol{p}$ was defined as the dipole moment per unit volume of a perovskite unit cell, taken to be centered at the B-site cation and bounded by its nearest Ba neighbors. 
All atoms belonging to the conventional cell were included, and their instantaneous positions were measured relative to the B-site reference position \cite{Sepliarsky_2011}:
\begin{equation}
\boldsymbol{p}=\frac{1}{v}\sum_i \frac{z_i}{w_i}\left(\boldsymbol{r}_i-\boldsymbol{r_{\rm B}}\right),
\label{SMpol}
\end{equation}
where $v$ is the unit-cell volume, $z_i$ and $\boldsymbol{r}_i$ are the charge and position of ion $i$, respectively, and $\boldsymbol{r_{\rm B}}$ denotes the position of the B-site atom. 
The factor $w_i$ accounts for the number of unit cells shared by atom $i$.

The MD simulations were performed using the LAMMPS (23 June 2022)  code \cite{LAMMPS} within a constant stress
and temperature (N, $\sigma$, T) ensemble. The in-plane stress components ($\sigma_{xx}$, $\sigma_{yy}$, $\sigma_{xy}$ )were controlled using a barostat and allowed to relax to the target pressure, while no barostat was applied along the out-of-plane direction. As a result, the out-of-plane lattice parameter was free to evolve during the simulation, mimicking a freestanding thin layer. A time step of 0.2 fs was used for the integration of the equations of motion. Temperature and pressure were controlled using Nosé-Hoover thermostats and barostats. The slabs were terminated with BaO planes on both surfaces, a choice motivated by their chemical stability and the fact that BaO-BaO interfaces exhibit weak, van der Waals–like interactions that are consistent with experimentally realized freestanding oxide membranes \cite{Cohen_1997,Lee_2024}. Periodic boundary conditions were imposed along the in-plane directions (x and y).
The systems were first equilibrated at 20 K and then heated up to 300 K to reach the paraelectric phase within the model. From this state, an annealing protocol was applied by cooling the system in temperature steps of 20 K, concluding with a final simulation at 5 K. At each temperature, the system was evolved for a total of $40 ps$, including an initial equilibration period followed by an additional $20 ps$ stage during which structural and polarization data were collected. This protocol was used to better sample stable configurations and avoid trapping the system in metastable states that can appear when simulations start directly from low temperatures. 

Because the atomistic BTO model underestimates the bulk Curie temperature (yielding $\approx$ 300 K compared to the experimental range of 390 K- 400 K), all temperatures were rescaled by a factor of 1.34, corresponding to the approximate ratio between these values. This factor is independent of the present simulations. The temperature values reported throughout the manuscript correspond to these rescaled temperatures, while the simulations themselves were performed using the unscaled temperature range described above. This temperature adjustment is widely used in atomistic-model studies \cite{Stachiotti_2011,Machado_2019,Nahas_2020,Zhang_2023}. Although the system studied here is not bulk, applying this correction provides a more meaningful reference scale and facilitates comparison with experimentally relevant temperature ranges. We note that quantum zero-point motion is not included in the present classical simulations and may become relevant at very low temperatures; however, the main conclusions rely on relative phase stability and finite-temperature trends, which are not expected to be qualitatively affected by this limitation.

To explore the stability of different phases, we considered cubic, tetragonal, orthorhombic and rhombohedral unit-cell starting configurations. To assess finite-size effects, we carried out simulations varying lateral dimensions and thicknesses at different temperatures. Thermal evolution and structural properties were then analyzed for each configuration.

Stress-free thin layers were modeled using simulation cells of size N$_x$$\times$N$_y$$\times$N$_z$, where N$_i$ (with $i = x, y, z$) denotes the number of Ti atoms along each pseudocubic direction. 

We analyze systems with in-plane dimensions ranging from N$_x$ = N$_y$ = 15 to 40 and thicknesses between N$_z$ = 1 to 20.

This systematic exploration of layers with different lateral sizes under periodic boundary conditions provides access to polarization configurations that represent simplified versions of those emerging in larger systems. At reduced scales, the system can stabilize a single, sometimes constrained, polarization state and allows the isolation of fundamental mechanisms. As the system size increases, these simple configurations evolve into more complex states involving multiple domains and dynamically interacting metastable structures. The size-dependent study therefore reveals how intricate polarization patterns emerge from basic structural units. Small systems expose the essential ingredients of collective behavior, whereas larger ones display emergent phenomena that arise only at extended scales. Identifying the stable configurations also provides valuable insight into the multiplicity of metastable states and lays the groundwork for understanding possible switching pathways between them.

To analyze the dynamics of the out-of-plane polarization, the instantaneous structure factor was calculated according~\cite{Ortiz_2024}
\begin{equation}
S(q_x,q_y,t)=\left|\sum_{x=0}^{d-1}\sum_{y=0}^{d-1} e^{-i2\pi(xq_x+yq_y)}\bar{P_z}(x,y,t)\right|^2
\end{equation}
 which corresponds to the Fourier transform of the out-of-plane polarization component $\bar{P_z}(x,y,t)$. We then average $S(\boldsymbol{q}, t)$ over the four central atomic planes and over a simulation time of 1600 ps for each temperature.

Streamlines were generated in ParaView \cite{paraview} by interpolating the discrete polarization field, providing a qualitative visualization of how the polarization organizes into continuous toroidal loops.

\section{Data availability}
Data underlying the figures and conclusions of this work are publicly available via Zenodo via~\href{https://doi.org/10.5281/zenodo.18327702}{10.5281/zenodo.18327702}.

\begin{suppinfo}

The Supporting Information is available free of charge at xxx
\begin{itemize}
  \item si.pdf contains additional details about the topological analysis, temperature evolution, structural analysis and flux-closure domains.
  \item movie\_s1.mp4 shows the dynamics of the switching between the chiral bubble and wave-helix phases under THz irradiation.
\end{itemize}
\end{suppinfo}

\begin{acknowledgement}
We thank M. G. Stachiotti, M. Sepliarsky, P. Márton, M. Paściak, M. Graf, and M. A. P. Gonçalves for insightful discussions and constructive criticism.
The authors acknowledge the support provided by the European Union by the ERC-STG project 2D-sandwich (Grant No 101040057) and by the Ferroic Multifunctionalities project, supported by the Ministry of Education, Youth, and Sports of the Czech Republic; Project No. CZ.02.01.01\slash00\slash22\_008\slash0004591, co-funded by the European Union. Computational resources were provided by the e-INFRA CZ project (ID:90254), supported by the Ministry of Education, Youth and Sports of the Czech Republic.
\end{acknowledgement}


\begin{mcitethebibliography}{48}
\providecommand*\natexlab[1]{#1}
\providecommand*\mciteSetBstSublistMode[1]{}
\providecommand*\mciteSetBstMaxWidthForm[2]{}
\providecommand*\mciteBstWouldAddEndPuncttrue
  {\def\EndOfBibitem{\unskip.}}
\providecommand*\mciteBstWouldAddEndPunctfalse
  {\let\EndOfBibitem\relax}
\providecommand*\mciteSetBstMidEndSepPunct[3]{}
\providecommand*\mciteSetBstSublistLabelBeginEnd[3]{}
\providecommand*\EndOfBibitem{}
\mciteSetBstSublistMode{f}
\mciteSetBstMaxWidthForm{subitem}{(\alph{mcitesubitemcount})}
\mciteSetBstSublistLabelBeginEnd
  {\mcitemaxwidthsubitemform\space}
  {\relax}
  {\relax}

\bibitem[Das \latin{et~al.}(2019)Das, Tang, Hong, Gon{\c{c}}alves, McCarter, Klewe, Nguyen, G{\'o}mez-Ortiz, Shafer, Arenholz, Stoica, Hsu, Wang, Ophus, Liu, Nelson, Saremi, Prasad, Mei, Schlom, {\'I}{\~{n}}iguez, Garc{\'i}a-Fern{\'a}ndez, Muller, Chen, Junquera, Martin, and Ramesh]{Das2019}
Das,~S.; Tang,~Y.~L.; Hong,~Z.; Gon{\c{c}}alves,~M. A.~P.; McCarter,~M.~R.; Klewe,~C.; Nguyen,~K.~X.; G{\'o}mez-Ortiz,~F.; Shafer,~P.; Arenholz,~E.; Stoica,~V.~A.; Hsu,~S.-L.; Wang,~B.; Ophus,~C.; Liu,~J.~F.; Nelson,~C.~T.; Saremi,~S.; Prasad,~B.; Mei,~A.~B.; Schlom,~D.~G.; {\'I}{\~{n}}iguez,~J.; Garc{\'i}a-Fern{\'a}ndez,~P.; Muller,~D.~A.; Chen,~L.~Q.; Junquera,~J.; Martin,~L.~W.; Ramesh,~R. Observation of room-temperature polar skyrmions. \emph{Nature} \textbf{2019}, \emph{568}, 368--372, DOI: \doi{10.1038/s41586-019-1092-8}\relax
\mciteBstWouldAddEndPuncttrue
\mciteSetBstMidEndSepPunct{\mcitedefaultmidpunct}
{\mcitedefaultendpunct}{\mcitedefaultseppunct}\relax
\EndOfBibitem
\bibitem[Das \latin{et~al.}(2020)Das, Hong, McCarter, Shafer, Shao, Muller, Martin, and Ramesh]{Das_2020}
Das,~S.; Hong,~Z.; McCarter,~M.; Shafer,~P.; Shao,~Y.-T.; Muller,~D.~A.; Martin,~L.~W.; Ramesh,~R. A new era in ferroelectrics. \emph{APL Mater.} \textbf{2020}, \emph{8}, 120902, DOI: \doi{10.1063/5.0034914}\relax
\mciteBstWouldAddEndPuncttrue
\mciteSetBstMidEndSepPunct{\mcitedefaultmidpunct}
{\mcitedefaultendpunct}{\mcitedefaultseppunct}\relax
\EndOfBibitem
\bibitem[Shao \latin{et~al.}(2023)Shao, Das, Hong, Xu, Chandrika, Gómez-Ortiz, García-Fernández, Chen, Hwang, Junquera, Martin, Ramesh, and Muller]{Shao2023}
Shao,~Y.-T.; Das,~S.; Hong,~Z.; Xu,~R.; Chandrika,~S.; Gómez-Ortiz,~F.; García-Fernández,~P.; Chen,~L.-Q.; Hwang,~H.~Y.; Junquera,~J.; Martin,~L.~W.; Ramesh,~R.; Muller,~D.~A. Emergent chirality in a polar meron to skyrmion phase transition. \emph{Nat. Commun.} \textbf{2023}, \emph{14}, 1355, DOI: \doi{10.1038/s41467-023-36950-x}\relax
\mciteBstWouldAddEndPuncttrue
\mciteSetBstMidEndSepPunct{\mcitedefaultmidpunct}
{\mcitedefaultendpunct}{\mcitedefaultseppunct}\relax
\EndOfBibitem
\bibitem[Junquera \latin{et~al.}(2023)Junquera, Nahas, Prokhorenko, Bellaiche, Iñiguez, Schlom, Chen, Salahuddin, Muller, Martin, and Ramesh]{Junquera_2023}
Junquera,~J.; Nahas,~Y.; Prokhorenko,~S.; Bellaiche,~L.; Iñiguez,~J.; Schlom,~D.~G.; Chen,~L.-Q.; Salahuddin,~S.; Muller,~D.~A.; Martin,~L.~W.; Ramesh,~R. Topological phases in polar oxide nanostructures. \emph{Rev. Mod. Phys.} \textbf{2023}, \emph{95}, 025001, DOI: \doi{10.1103/RevModPhys.95.025001}\relax
\mciteBstWouldAddEndPuncttrue
\mciteSetBstMidEndSepPunct{\mcitedefaultmidpunct}
{\mcitedefaultendpunct}{\mcitedefaultseppunct}\relax
\EndOfBibitem
\bibitem[Lukyanchuk \latin{et~al.}(2025)Lukyanchuk, Razumnaya, Kondovych, Tikhonov, Khesin, and Vinokur]{Lukyanchuk_2025}
Lukyanchuk,~I.~A.; Razumnaya,~A.~G.; Kondovych,~S.; Tikhonov,~Y.~A.; Khesin,~B.; Vinokur,~V.~M. Topological foundations of ferroelectricity. \emph{Phys. Rep.} \textbf{2025}, \emph{1110}, 1--56, DOI: \doi{10.1016/j.physrep.2025.01.002}\relax
\mciteBstWouldAddEndPuncttrue
\mciteSetBstMidEndSepPunct{\mcitedefaultmidpunct}
{\mcitedefaultendpunct}{\mcitedefaultseppunct}\relax
\EndOfBibitem
\bibitem[Catalan \latin{et~al.}(2012)Catalan, Seidel, Ramesh, and Scott]{Catalan_2012}
Catalan,~G.; Seidel,~J.; Ramesh,~R.; Scott,~J.~F. Domain wall nanoelectronics. \emph{Rev. Mod. Phys.} \textbf{2012}, \emph{84}, 119--156, DOI: \doi{10.1103/RevModPhys.84.119}\relax
\mciteBstWouldAddEndPuncttrue
\mciteSetBstMidEndSepPunct{\mcitedefaultmidpunct}
{\mcitedefaultendpunct}{\mcitedefaultseppunct}\relax
\EndOfBibitem
\bibitem[Chen \latin{et~al.}(2020)Chen, Yuan, Hou, Tang, Zhang, Wang, Li, Zhao, Liu, Chen, Martin, and Chen]{Chen_2020}
Chen,~S.; Yuan,~S.; Hou,~Z.; Tang,~Y.; Zhang,~J.; Wang,~T.; Li,~K.; Zhao,~W.; Liu,~X.; Chen,~L.; Martin,~L.~W.; Chen,~Z. Recent Progress on Topological Structures in Ferroic Thin Films and Heterostructures. \emph{Adv. Mater.} \textbf{2020}, \emph{33}, 2000857, DOI: \doi{10.1002/adma.202000857}\relax
\mciteBstWouldAddEndPuncttrue
\mciteSetBstMidEndSepPunct{\mcitedefaultmidpunct}
{\mcitedefaultendpunct}{\mcitedefaultseppunct}\relax
\EndOfBibitem
\bibitem[Hu \latin{et~al.}(2024)Hu, Yang, and Liu]{Hu_2024}
Hu,~Y.; Yang,~J.; Liu,~S. Giant Piezoelectric Effects of Topological Structures in Stretched Ferroelectric Membranes. \emph{Phys. Rev. Lett.} \textbf{2024}, \emph{133}, 046802, DOI: \doi{10.1103/PhysRevLett.133.046802}\relax
\mciteBstWouldAddEndPuncttrue
\mciteSetBstMidEndSepPunct{\mcitedefaultmidpunct}
{\mcitedefaultendpunct}{\mcitedefaultseppunct}\relax
\EndOfBibitem
\bibitem[Rabe(2005)]{Rabe_2005}
Rabe,~K.~M. Theoretical investigations of epitaxial strain effects in ferroelectric oxide thin films and superlattices. \emph{Curr. Opin. Solid State Mater. Sci.} \textbf{2005}, \emph{9}, 122--127, DOI: \doi{10.1016/j.cossms.2006.06.003}\relax
\mciteBstWouldAddEndPuncttrue
\mciteSetBstMidEndSepPunct{\mcitedefaultmidpunct}
{\mcitedefaultendpunct}{\mcitedefaultseppunct}\relax
\EndOfBibitem
\bibitem[Tinte and Stachiotti(2001)Tinte, and Stachiotti]{Tinte_2001}
Tinte,~S.; Stachiotti,~M. Surface effects and ferroelectric phase transitions in {BaTiO$_3$} ultrathin films. \emph{Phys. Rev. B} \textbf{2001}, \emph{64}, 235403, DOI: \doi{10.1103/PhysRevB.64.235403}\relax
\mciteBstWouldAddEndPuncttrue
\mciteSetBstMidEndSepPunct{\mcitedefaultmidpunct}
{\mcitedefaultendpunct}{\mcitedefaultseppunct}\relax
\EndOfBibitem
\bibitem[Pertsev \latin{et~al.}(1998)Pertsev, Zembilgotov, and Tagantsev]{Pertsev_1988}
Pertsev,~N.~A.; Zembilgotov,~A.~G.; Tagantsev,~A.~K. Effect of mechanical boundary conditions on phase diagrams of epitaxial ferroelectric thin films. \emph{Phys. Rev. Lett.} \textbf{1998}, \emph{80}, 1988--1991, DOI: \doi{10.1103/PhysRevLett.80.1988}\relax
\mciteBstWouldAddEndPuncttrue
\mciteSetBstMidEndSepPunct{\mcitedefaultmidpunct}
{\mcitedefaultendpunct}{\mcitedefaultseppunct}\relax
\EndOfBibitem
\bibitem[Pertsev \latin{et~al.}(1999)Pertsev, Zembilgotov, and Tagantsev]{Pertsev_1999}
Pertsev,~N.~A.; Zembilgotov,~A.~G.; Tagantsev,~A.~K. Equilibrium states and phase transitions in epitaxial ferroelectric thin films. \emph{Ferroelectrics} \textbf{1999}, \emph{223}, 79--90, DOI: \doi{10.1080/00150199908260556}\relax
\mciteBstWouldAddEndPuncttrue
\mciteSetBstMidEndSepPunct{\mcitedefaultmidpunct}
{\mcitedefaultendpunct}{\mcitedefaultseppunct}\relax
\EndOfBibitem
\bibitem[Wu and Li(2021)Wu, and Li]{Wu_2021}
Wu,~M.; Li,~J. Sliding ferroelectricity in {2D} van der waals materials: related physics and future opportunities. \emph{Proc. Natl. Acad. Sci. U. S. A.} \textbf{2021}, \emph{118}, e2115703118, DOI: \doi{10.1073/pnas.2115703118}\relax
\mciteBstWouldAddEndPuncttrue
\mciteSetBstMidEndSepPunct{\mcitedefaultmidpunct}
{\mcitedefaultendpunct}{\mcitedefaultseppunct}\relax
\EndOfBibitem
\bibitem[Weston \latin{et~al.}(2022)Weston, Castanon, Enaldiev, Ferreira, Bhattacharjee, Xu, Corte-Le{\'o}n, Wu, Clark, Summerfield, Hashimoto, Gao, Wang, Hamer, Read, Fumagalli, Kretinin, Haigh, Kazakova, Geim, Fal'ko, and Gorbachev]{Weston_2022}
Weston,~A.; Castanon,~E.~G.; Enaldiev,~V.; Ferreira,~F.; Bhattacharjee,~S.; Xu,~S.; Corte-Le{\'o}n,~H.; Wu,~Z.; Clark,~N.; Summerfield,~A.; Hashimoto,~T.; Gao,~Y.; Wang,~W.; Hamer,~M.; Read,~H.; Fumagalli,~L.; Kretinin,~A.~V.; Haigh,~S.~J.; Kazakova,~O.; Geim,~A.~K.; Fal'ko,~V.~I.; Gorbachev,~R. Interfacial ferroelectricity in marginally twisted {2D} semiconductors. \emph{Nature Nanotechnology} \textbf{2022}, \emph{17}, 390--395, DOI: \doi{10.1038/s41565-022-01072-w}\relax
\mciteBstWouldAddEndPuncttrue
\mciteSetBstMidEndSepPunct{\mcitedefaultmidpunct}
{\mcitedefaultendpunct}{\mcitedefaultseppunct}\relax
\EndOfBibitem
\bibitem[Hassan \latin{et~al.}(2024)Hassan, Singh, Joe, Son, Ngo, Jang, Sett, Singha, Biswas, Bhakar, Watanabe, Taniguchi, Raghunathan, Sheet, Lee, Yoo, Srivastava, and Lee]{Hassan_2024}
Hassan,~Y.; Singh,~B.; Joe,~M.; Son,~B.-M.; Ngo,~T.~D.; Jang,~Y.; Sett,~S.; Singha,~A.; Biswas,~R.; Bhakar,~M.; Watanabe,~K.; Taniguchi,~T.; Raghunathan,~V.; Sheet,~G.; Lee,~Z.; Yoo,~W.~J.; Srivastava,~P.~K.; Lee,~C. Twist-controlled ferroelectricity and emergent multiferroicity in {WSe$_2$} bilayers. \emph{Adv. Mater.} \textbf{2024}, \emph{36}, 2406290, DOI: \doi{10.1002/adma.202406290}\relax
\mciteBstWouldAddEndPuncttrue
\mciteSetBstMidEndSepPunct{\mcitedefaultmidpunct}
{\mcitedefaultendpunct}{\mcitedefaultseppunct}\relax
\EndOfBibitem
\bibitem[Li \latin{et~al.}(2024)Li, Wang, Wang, Yang, Wang, Zhan, He, and Wang]{Li_2024}
Li,~S.; Wang,~F.; Wang,~Y.; Yang,~J.; Wang,~X.; Zhan,~X.; He,~J.; Wang,~Z. Van der {W}aals ferroelectrics: theories, materials, and device applications. \emph{Adv. Mater.} \textbf{2024}, \emph{36}, 2301472, DOI: \doi{10.1002/adma.202301472}\relax
\mciteBstWouldAddEndPuncttrue
\mciteSetBstMidEndSepPunct{\mcitedefaultmidpunct}
{\mcitedefaultendpunct}{\mcitedefaultseppunct}\relax
\EndOfBibitem
\bibitem[Novoselov \latin{et~al.}(2004)Novoselov, Geim, Morozov, Jiang, Zhang, Dubonos, Grigorieva, and Firsov]{Novoselov2004}
Novoselov,~K.~S.; Geim,~A.~K.; Morozov,~S.~V.; Jiang,~D.; Zhang,~Y.; Dubonos,~S.~V.; Grigorieva,~I.~V.; Firsov,~A.~A. Electric Field Effect in Atomically Thin Carbon Films. \emph{Science} \textbf{2004}, \emph{306}, 666--669, DOI: \doi{10.1126/science.1102896}\relax
\mciteBstWouldAddEndPuncttrue
\mciteSetBstMidEndSepPunct{\mcitedefaultmidpunct}
{\mcitedefaultendpunct}{\mcitedefaultseppunct}\relax
\EndOfBibitem
\bibitem[Castro~Neto \latin{et~al.}(2009)Castro~Neto, Guinea, Peres, Novoselov, and Geim]{Castro_2009}
Castro~Neto,~A.~H.; Guinea,~F.; Peres,~N.~M.; Novoselov,~K.~S.; Geim,~A.~K. The electronic properties of graphene. \emph{Rev. Mod. Phys.} \textbf{2009}, \emph{81}, 109--162, DOI: \doi{10.1103/RevModPhys.81.109}\relax
\mciteBstWouldAddEndPuncttrue
\mciteSetBstMidEndSepPunct{\mcitedefaultmidpunct}
{\mcitedefaultendpunct}{\mcitedefaultseppunct}\relax
\EndOfBibitem
\bibitem[Fernandez \latin{et~al.}(2022)Fernandez, Acharya, Lee, Schimpf, Jiang, Lou, Tian, and Martin]{Fernandez_2022}
Fernandez,~A.; Acharya,~M.; Lee,~H.-G.; Schimpf,~J.; Jiang,~Y.; Lou,~D.; Tian,~Z.; Martin,~L.~W. Thin-Film Ferroelectrics. \emph{Adv. Mater.} \textbf{2022}, \emph{34}, 2108841, DOI: \doi{10.1002/adma.202108841}\relax
\mciteBstWouldAddEndPuncttrue
\mciteSetBstMidEndSepPunct{\mcitedefaultmidpunct}
{\mcitedefaultendpunct}{\mcitedefaultseppunct}\relax
\EndOfBibitem
\bibitem[Chiabrera \latin{et~al.}(2022)Chiabrera, Yun, Li, Dahm, Zhang, Kirchert, Christensen, Trier, Jespersen, and Pryds]{Chiabrera_2022}
Chiabrera,~F.~M.; Yun,~S.; Li,~Y.; Dahm,~R.~T.; Zhang,~H.; Kirchert,~C. K.~R.; Christensen,~D.~V.; Trier,~F.; Jespersen,~T.~S.; Pryds,~N. Freestanding perovskite oxide films: synthesis, challenges, and properties. \emph{Ann. Phys.} \textbf{2022}, \emph{534}, 2200084, DOI: \doi{10.1002/andp.202200084}\relax
\mciteBstWouldAddEndPuncttrue
\mciteSetBstMidEndSepPunct{\mcitedefaultmidpunct}
{\mcitedefaultendpunct}{\mcitedefaultseppunct}\relax
\EndOfBibitem
\bibitem[S{\'a}nchez-Santolino \latin{et~al.}(2024)S{\'a}nchez-Santolino, Rouco, Puebla, Aramberri, Zamora, Cabero, Cuellar, Munuera, Mompean, Garcia-Hernandez, Castellanos-Gomez, {\'I}{\~{n}}iguez, Leon, and Santamaria]{Sanchez_2024}
S{\'a}nchez-Santolino,~G.; Rouco,~V.; Puebla,~S.; Aramberri,~H.; Zamora,~V.; Cabero,~M.; Cuellar,~F.~A.; Munuera,~C.; Mompean,~F.; Garcia-Hernandez,~M.; Castellanos-Gomez,~A.; {\'I}{\~{n}}iguez,~J.; Leon,~C.; Santamaria,~J. A {2D} ferroelectric vortex pattern in twisted {BaTiO$_3$} freestanding layers. \emph{Nature} \textbf{2024}, \emph{626}, 529--534, DOI: \doi{10.1038/s41586-023-06978-6}\relax
\mciteBstWouldAddEndPuncttrue
\mciteSetBstMidEndSepPunct{\mcitedefaultmidpunct}
{\mcitedefaultendpunct}{\mcitedefaultseppunct}\relax
\EndOfBibitem
\bibitem[Sha \latin{et~al.}(2024)Sha, Zhang, Ma, Li, Yang, Cui, Li, Huang, and Yu]{Sha_2024}
Sha,~H.; Zhang,~Y.; Ma,~Y.; Li,~W.; Yang,~W.; Cui,~J.; Li,~Q.; Huang,~H.; Yu,~R. Polar vortex hidden in twisted bilayers of paraelectric {SrTiO$_3$}. \emph{Nat. Commun.} \textbf{2024}, \emph{15}, 10915, DOI: \doi{10.1038/s41467-024-55328-1}\relax
\mciteBstWouldAddEndPuncttrue
\mciteSetBstMidEndSepPunct{\mcitedefaultmidpunct}
{\mcitedefaultendpunct}{\mcitedefaultseppunct}\relax
\EndOfBibitem
\bibitem[Zhang \latin{et~al.}(2024)Zhang, Zhang, Cui, and Zhang]{Zhang_2024}
Zhang,~C.; Zhang,~S.; Cui,~P.; Zhang,~Z. Tunable multistate ferroelectricity of unit-cell-thick {BaTiO$_3$} revived by a ferroelectric {SnS} monolayer via interfacial sliding. \emph{Nano Lett.} \textbf{2024}, \emph{24}, 8664--8670, DOI: \doi{10.1021/acs.nanolett.4c02041}\relax
\mciteBstWouldAddEndPuncttrue
\mciteSetBstMidEndSepPunct{\mcitedefaultmidpunct}
{\mcitedefaultendpunct}{\mcitedefaultseppunct}\relax
\EndOfBibitem
\bibitem[Lee \latin{et~al.}(2024)Lee, de~Sousa, Jalan, and Low]{Lee_2024}
Lee,~S.; de~Sousa,~D. J.~P.; Jalan,~B.; Low,~T. Moiré polar vortex, flat bands, and Lieb lattice in twisted bilayer {BaTiO$_3$}. \emph{Sci. Adv.} \textbf{2024}, \emph{10}, eadq0293, DOI: \doi{10.1126/sciadv.adq0293}\relax
\mciteBstWouldAddEndPuncttrue
\mciteSetBstMidEndSepPunct{\mcitedefaultmidpunct}
{\mcitedefaultendpunct}{\mcitedefaultseppunct}\relax
\EndOfBibitem
\bibitem[Zubko \latin{et~al.}(2016)Zubko, Wojde{\l}, Hadjimichael, Fernandez-Pena, Sen{\'e}, Luk’yanchuk, Triscone, and {\'I}{\~n}iguez]{Zubko_2016}
Zubko,~P.; Wojde{\l},~J.~C.; Hadjimichael,~M.; Fernandez-Pena,~S.; Sen{\'e},~A.; Luk’yanchuk,~I.; Triscone,~J.-M.; {\'I}{\~n}iguez,~J. Negative capacitance in multidomain ferroelectric superlattices. \emph{Nature} \textbf{2016}, \emph{534}, 524--528, DOI: \doi{10.1038/nature17659}\relax
\mciteBstWouldAddEndPuncttrue
\mciteSetBstMidEndSepPunct{\mcitedefaultmidpunct}
{\mcitedefaultendpunct}{\mcitedefaultseppunct}\relax
\EndOfBibitem
\bibitem[G\'omez-Ortiz \latin{et~al.}(2024)G\'omez-Ortiz, Graf, Junquera, Iñiguez-Gonz\'alez, and Aramberri]{Ortiz_2024}
G\'omez-Ortiz,~F.; Graf,~M.; Junquera,~J.; Iñiguez-Gonz\'alez,~J.; Aramberri,~H. Liquid-crystal-like dynamic transition in ferroelectric-dielectric superlattices. \emph{Phys. Rev. Lett.} \textbf{2024}, \emph{133}, 066801, DOI: \doi{10.1103/PhysRevLett.133.066801}\relax
\mciteBstWouldAddEndPuncttrue
\mciteSetBstMidEndSepPunct{\mcitedefaultmidpunct}
{\mcitedefaultendpunct}{\mcitedefaultseppunct}\relax
\EndOfBibitem
\bibitem[Nahas \latin{et~al.}(2020)Nahas, Prokhorenko, Zhang, Govinden, Valanoor, and Bellaiche]{Nahas_2020}
Nahas,~Y.; Prokhorenko,~S.; Zhang,~Q.; Govinden,~V.; Valanoor,~N.; Bellaiche,~L. Topology and control of self-assembled domain patterns in low-dimensional ferroelectrics. \emph{Nat. Commun.} \textbf{2020}, \emph{11}, 5779, DOI: \doi{10.1038/s41467-020-19519-w}\relax
\mciteBstWouldAddEndPuncttrue
\mciteSetBstMidEndSepPunct{\mcitedefaultmidpunct}
{\mcitedefaultendpunct}{\mcitedefaultseppunct}\relax
\EndOfBibitem
\bibitem[Nahas \latin{et~al.}(2020)Nahas, Prokhorenko, Fischer, Xu, Carr{\'e}t{\'e}ro, Prosandeev, Bibes, Fusil, Dkhil, Garcia, and Bellaiche]{Nahas_2020_2}
Nahas,~Y.; Prokhorenko,~S.; Fischer,~J.; Xu,~B.; Carr{\'e}t{\'e}ro,~C.; Prosandeev,~S.; Bibes,~M.; Fusil,~S.; Dkhil,~B.; Garcia,~V.; Bellaiche,~L. Inverse transition of labyrinthine domain patterns in ferroelectric thin films. \emph{Nature} \textbf{2020}, \emph{577}, 47--51, DOI: \doi{10.1038/s41586-019-1845-4}\relax
\mciteBstWouldAddEndPuncttrue
\mciteSetBstMidEndSepPunct{\mcitedefaultmidpunct}
{\mcitedefaultendpunct}{\mcitedefaultseppunct}\relax
\EndOfBibitem
\bibitem[Boron \latin{et~al.}(2025)Boron, Sené, Tikhonov, Razumnaya, Lukyanchuk, and Kondovych]{Boron_2025}
Boron,~L.; Sené,~A.; Tikhonov,~Y.; Razumnaya,~A.; Lukyanchuk,~I.; Kondovych,~S. Morphology of polarization states in strained ferroelectric films.  arXiv (cond-mat.mtrl-sci), submitted 2025-09-08, DOI: \doi{10.48550/arXiv.2509.06508},  (accessed 2025–12–01)\relax
\mciteBstWouldAddEndPuncttrue
\mciteSetBstMidEndSepPunct{\mcitedefaultmidpunct}
{\mcitedefaultendpunct}{\mcitedefaultseppunct}\relax
\EndOfBibitem
\bibitem[Kondovych \latin{et~al.}(2025)Kondovych, Boron, Di~Rino, Sepliarsky, Razumnaya, Sen{\'e}, and Lukyanchuk]{Kondovych_2025}
Kondovych,~S.; Boron,~L.; Di~Rino,~F.~N.; Sepliarsky,~M.; Razumnaya,~A.~G.; Sen{\'e},~A.; Lukyanchuk,~I.~A. Surface-tension-induced phase transitions in freestanding ferroelectric thin films. \emph{Nano Lett.} \textbf{2025}, \emph{25}, 12987--12994, DOI: \doi{10.1021/acs.nanolett.5c03216}\relax
\mciteBstWouldAddEndPuncttrue
\mciteSetBstMidEndSepPunct{\mcitedefaultmidpunct}
{\mcitedefaultendpunct}{\mcitedefaultseppunct}\relax
\EndOfBibitem
\bibitem[Marton \latin{et~al.}(2010)Marton, Rychetsky, and Hlinka]{Marton2010}
Marton,~P.; Rychetsky,~I.; Hlinka,~J. Domain walls of ferroelectric {BaTiO$_3$} within the Ginzburg-Landau-Devonshire phenomenological model. \emph{Phys. Rev. B} \textbf{2010}, \emph{81}, 144125, DOI: \doi{10.1103/PhysRevB.81.144125}\relax
\mciteBstWouldAddEndPuncttrue
\mciteSetBstMidEndSepPunct{\mcitedefaultmidpunct}
{\mcitedefaultendpunct}{\mcitedefaultseppunct}\relax
\EndOfBibitem
\bibitem[Nahas \latin{et~al.}(2015)Nahas, Prokhorenko, Louis, Gui, Kornev, and Bellaiche]{Nahas_2015}
Nahas,~Y.; Prokhorenko,~S.; Louis,~L.; Gui,~Z.; Kornev,~I.; Bellaiche,~L. Discovery of stable skyrmionic state in ferroelectric nanocomposites. \emph{Nat. Commun.} \textbf{2015}, \emph{6}, 8542, DOI: \doi{10.1038/ncomms9542}\relax
\mciteBstWouldAddEndPuncttrue
\mciteSetBstMidEndSepPunct{\mcitedefaultmidpunct}
{\mcitedefaultendpunct}{\mcitedefaultseppunct}\relax
\EndOfBibitem
\bibitem[Gon{\c c}alves \latin{et~al.}(2024)Gon{\c c}alves, Pa{\'s}ciak, and Hlinka]{Gonsalves_2024}
Gon{\c c}alves,~M. A.~P.; Pa{\'s}ciak,~M.; Hlinka,~J. Antiskyrmions in Ferroelectric Barium Titanate. \emph{Phys. Rev. Lett.} \textbf{2024}, \emph{133}, 066802, DOI: \doi{10.1103/PhysRevLett.133.066802}\relax
\mciteBstWouldAddEndPuncttrue
\mciteSetBstMidEndSepPunct{\mcitedefaultmidpunct}
{\mcitedefaultendpunct}{\mcitedefaultseppunct}\relax
\EndOfBibitem
\bibitem[Bastogne \latin{et~al.}(2024)Bastogne, Gómez-Ortiz, Anand, and Ghosez]{Bastogne2024}
Bastogne,~L.; Gómez-Ortiz,~F.; Anand,~S.; Ghosez,~P. Dynamical manipulation of polar topologies from acoustic phonon excitations. \emph{Nano Lett.} \textbf{2024}, \emph{24}, 13783--13789, DOI: \doi{10.1021/acs.nanolett.4c04125}\relax
\mciteBstWouldAddEndPuncttrue
\mciteSetBstMidEndSepPunct{\mcitedefaultmidpunct}
{\mcitedefaultendpunct}{\mcitedefaultseppunct}\relax
\EndOfBibitem
\bibitem[Abutbul and Podolsky(2022)Abutbul, and Podolsky]{Abutbul_2022}
Abutbul,~D.; Podolsky,~D. Topological Order in an Antiferromagnetic Tetratic Phase. \emph{Phys. Rev. Lett.} \textbf{2022}, \emph{128}, 255501, DOI: \doi{10.1103/PhysRevLett.128.255501}\relax
\mciteBstWouldAddEndPuncttrue
\mciteSetBstMidEndSepPunct{\mcitedefaultmidpunct}
{\mcitedefaultendpunct}{\mcitedefaultseppunct}\relax
\EndOfBibitem
\bibitem[Sepliarsky \latin{et~al.}(2005)Sepliarsky, Asthagiri, Phillpot, Stachiotti, and Migoni]{Sepliarsky_2005}
Sepliarsky,~M.; Asthagiri,~A.; Phillpot,~S.; Stachiotti,~M.; Migoni,~R. Atomic-level simulation of ferroelectricity in oxide materials. \emph{Curr. Opin. Solid State Mater. Sci.} \textbf{2005}, \emph{9}, 107--113, DOI: \doi{https://doi.org/10.1016/j.cossms.2006.05.002}\relax
\mciteBstWouldAddEndPuncttrue
\mciteSetBstMidEndSepPunct{\mcitedefaultmidpunct}
{\mcitedefaultendpunct}{\mcitedefaultseppunct}\relax
\EndOfBibitem
\bibitem[Ghosez and Junquera(2022)Ghosez, and Junquera]{Ghosez_2022}
Ghosez,~P.; Junquera,~J. Modeling of ferroelectric oxide perovskites: from first to second principles. \emph{Annu. Rev. Condens. Matter Phys.} \textbf{2022}, \emph{13}, 325--364, DOI: \doi{10.1146/annurev-conmatphys-040220-045528}\relax
\mciteBstWouldAddEndPuncttrue
\mciteSetBstMidEndSepPunct{\mcitedefaultmidpunct}
{\mcitedefaultendpunct}{\mcitedefaultseppunct}\relax
\EndOfBibitem
\bibitem[Machado \latin{et~al.}(2019)Machado, {Di Loreto}, Frattini, Sepliarsky, and Stachiotti]{Machado_2019}
Machado,~R.; {Di Loreto},~A.; Frattini,~A.; Sepliarsky,~M.; Stachiotti,~M. Site occupancy effects of mg impurities in {BaTiO$_3$}. \emph{J. Alloys Compd.} \textbf{2019}, \emph{809}, 151847, DOI: \doi{https://doi.org/10.1016/j.jallcom.2019.151847}\relax
\mciteBstWouldAddEndPuncttrue
\mciteSetBstMidEndSepPunct{\mcitedefaultmidpunct}
{\mcitedefaultendpunct}{\mcitedefaultseppunct}\relax
\EndOfBibitem
\bibitem[Sepliarsky \latin{et~al.}(2023)Sepliarsky, Machado, Tinte, and Stachiotti]{Sepliarsky_2023}
Sepliarsky,~M.; Machado,~R.; Tinte,~S.; Stachiotti,~M.~G. Effects of the antiferrodistortive instability on the structural behavior of {BaZrO$_3$} by atomistic simulations. \emph{Phys. Rev. B} \textbf{2023}, \emph{107}, 134102, DOI: \doi{10.1103/PhysRevB.107.134102}\relax
\mciteBstWouldAddEndPuncttrue
\mciteSetBstMidEndSepPunct{\mcitedefaultmidpunct}
{\mcitedefaultendpunct}{\mcitedefaultseppunct}\relax
\EndOfBibitem
\bibitem[Sepliarsky \latin{et~al.}(2025)Sepliarsky, Aquistapace, Di~Rino, Machado, and Stachiotti]{Sepliarsky_2025}
Sepliarsky,~M.; Aquistapace,~F.; Di~Rino,~F.; Machado,~R.; Stachiotti,~M.~G. Local-phase framework for the {BaTi$_{1-x}$Zr$_x$O$_3$} phase diagram: From ferroelectricity to dipolar glass. \emph{Phys. Rev. B} \textbf{2025}, \emph{112}, 214105, DOI: \doi{10.1103/kz7p-t99d}\relax
\mciteBstWouldAddEndPuncttrue
\mciteSetBstMidEndSepPunct{\mcitedefaultmidpunct}
{\mcitedefaultendpunct}{\mcitedefaultseppunct}\relax
\EndOfBibitem
\bibitem[Stachiotti and Sepliarsky(2011)Stachiotti, and Sepliarsky]{Stachiotti_2011}
Stachiotti,~M.~G.; Sepliarsky,~M. Toroidal Ferroelectricity in {PbTiO$_3$} Nanoparticles. \emph{Phys. Rev. Lett.} \textbf{2011}, \emph{106}, 137601, DOI: \doi{10.1103/PhysRevLett.106.137601}\relax
\mciteBstWouldAddEndPuncttrue
\mciteSetBstMidEndSepPunct{\mcitedefaultmidpunct}
{\mcitedefaultendpunct}{\mcitedefaultseppunct}\relax
\EndOfBibitem
\bibitem[Di~Rino \latin{et~al.}(2025)Di~Rino, Boron, Pavlenko, Tikhonov, Razumnaya, Sepliarsky, Sen{\'e}, Lukyanchuk, and Kondovych]{DiRino_2025}
Di~Rino,~F.; Boron,~L.; Pavlenko,~M.~A.; Tikhonov,~I.~A.; Razumnaya,~A.~G.; Sepliarsky,~M.; Sen{\'e},~A.; Lukyanchuk,~I.~A.; Kondovych,~S. Topological states in ferroelectric nanorods tuned by the surface tension. \emph{Commun. Mater.} \textbf{2025}, \emph{6}, 57, DOI: \doi{10.1038/s43246-025-00774-7}\relax
\mciteBstWouldAddEndPuncttrue
\mciteSetBstMidEndSepPunct{\mcitedefaultmidpunct}
{\mcitedefaultendpunct}{\mcitedefaultseppunct}\relax
\EndOfBibitem
\bibitem[Sepliarsky and Cohen(2011)Sepliarsky, and Cohen]{Sepliarsky_2011}
Sepliarsky,~M.; Cohen,~R. First-principles based atomistic modeling of phase stability in {PMN-xPT}. \emph{J. Condens. Matter Phys.} \textbf{2011}, \emph{23}, 435902, DOI: \doi{10.1088/0953-8984/23/43/435902}\relax
\mciteBstWouldAddEndPuncttrue
\mciteSetBstMidEndSepPunct{\mcitedefaultmidpunct}
{\mcitedefaultendpunct}{\mcitedefaultseppunct}\relax
\EndOfBibitem
\bibitem[Thompson \latin{et~al.}(2022)Thompson, Aktulga, Berger, Bolintineanu, Brown, Crozier, in~'t Veld, Kohlmeyer, Moore, Nguyen, Shan, Stevens, Tranchida, Trott, and Plimpton]{LAMMPS}
Thompson,~A.~P.; Aktulga,~H.~M.; Berger,~R.; Bolintineanu,~D.~S.; Brown,~W.~M.; Crozier,~P.~S.; in~'t Veld,~P.~J.; Kohlmeyer,~A.; Moore,~S.~G.; Nguyen,~T.~D.; Shan,~R.; Stevens,~M.~J.; Tranchida,~J.; Trott,~C.; Plimpton,~S.~J. {LAMMPS} - a flexible simulation tool for particle-based materials modeling at the atomic, meso, and continuum scales. \emph{Comp. Phys. Comm.} \textbf{2022}, \emph{271}, 108171, DOI: \doi{10.1016/j.cpc.2021.108171}\relax
\mciteBstWouldAddEndPuncttrue
\mciteSetBstMidEndSepPunct{\mcitedefaultmidpunct}
{\mcitedefaultendpunct}{\mcitedefaultseppunct}\relax
\EndOfBibitem
\bibitem[Cohen(1997)]{Cohen_1997}
Cohen,~R.~E. Surface effects in ferroelectrics: Periodic slab computations for BaTiO$_3$. \emph{Ferroelectrics} \textbf{1997}, \emph{194}, 323--342, DOI: \doi{10.1080/00150199708016102}\relax
\mciteBstWouldAddEndPuncttrue
\mciteSetBstMidEndSepPunct{\mcitedefaultmidpunct}
{\mcitedefaultendpunct}{\mcitedefaultseppunct}\relax
\EndOfBibitem
\bibitem[Zhang \latin{et~al.}(2023)Zhang, Bastogne, He, Tang, Zhang, Ghosez, and Wang]{Zhang_2023}
Zhang,~J.; Bastogne,~L.; He,~X.; Tang,~G.; Zhang,~Y.; Ghosez,~P.; Wang,~J. Structural phase transitions and dielectric properties of {BaTiO$_3$} from a second-principles method. \emph{Phys. Rev. B} \textbf{2023}, \emph{108}, 134117, DOI: \doi{10.1103/PhysRevB.108.134117}\relax
\mciteBstWouldAddEndPuncttrue
\mciteSetBstMidEndSepPunct{\mcitedefaultmidpunct}
{\mcitedefaultendpunct}{\mcitedefaultseppunct}\relax
\EndOfBibitem
\bibitem[Ahrens \latin{et~al.}(2005)Ahrens, Geveci, and Law]{paraview}
Ahrens,~J.; Geveci,~B.; Law,~C. In \emph{Visualization Handbook}; Johnson,~C.~R., Hansen,~C.~D., Eds.; Elsevier, 2005\relax
\mciteBstWouldAddEndPuncttrue
\mciteSetBstMidEndSepPunct{\mcitedefaultmidpunct}
{\mcitedefaultendpunct}{\mcitedefaultseppunct}\relax
\EndOfBibitem
\end{mcitethebibliography}
\providecommand{\latin}[1]{#1}
\makeatletter
\providecommand{\doi}
  {\begingroup\let\do\@makeother\dospecials
  \catcode`\{=1 \catcode`\}=2 \doi@aux}
\providecommand{\doi@aux}[1]{\endgroup\texttt{#1}}
\makeatother
\providecommand*\mcitethebibliography{\thebibliography}
\csname @ifundefined\endcsname{endmcitethebibliography}  {\let\endmcitethebibliography\endthebibliography}{}

\end{document}





\section{Topological analysis}

To characterize the chiral bubble–domain structure, we computed the vorticity $\boldsymbol{\omega}$ of the polarization field $\boldsymbol{P}$ \cite{Lukyanchuk_2025},

\begin{equation}
\boldsymbol{\omega} = \nabla \times \boldsymbol{P},
\end{equation}

and identified helical cores by thresholding regions with large vorticity magnitude.

Streamlines were generated in ParaView \cite{paraview} by interpolating the discrete polarization field, providing a qualitative visualization of how the polarization organizes into continuous toroidal loops. This representation highlights the three-dimensional arrangement of the helical cores, which periodically migrate between the top and bottom film surfaces. The analysis is intended as a qualitative tool to reveal the underlying topology rather than as a quantitative reconstruction of the vector field.

\begin{figure}[ht]
\centering
\includegraphics[width=0.56\linewidth]{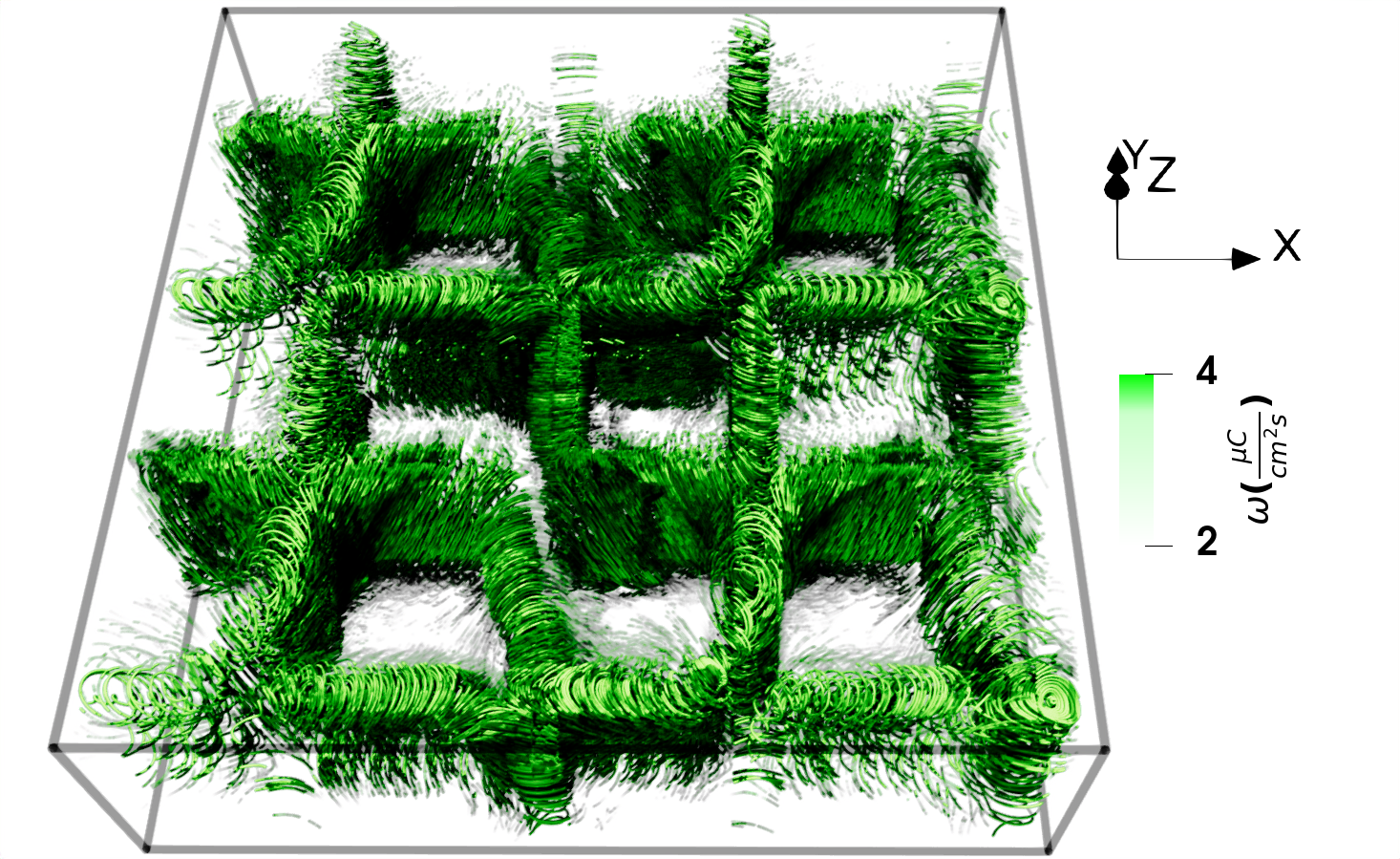}
\caption{\label{fig:core} Thresholded vorticity regions. Streamlines highlight the local polarization flow and reveal how the core axes gradually tilt and bend, eventually tending to close into loop-like structures.}
\end{figure}

\section{Temperature evolution}
\begin{figure*}[ht]
\centering
\vspace{0.1cm}
\begin{overpic}[width=0.47\linewidth]{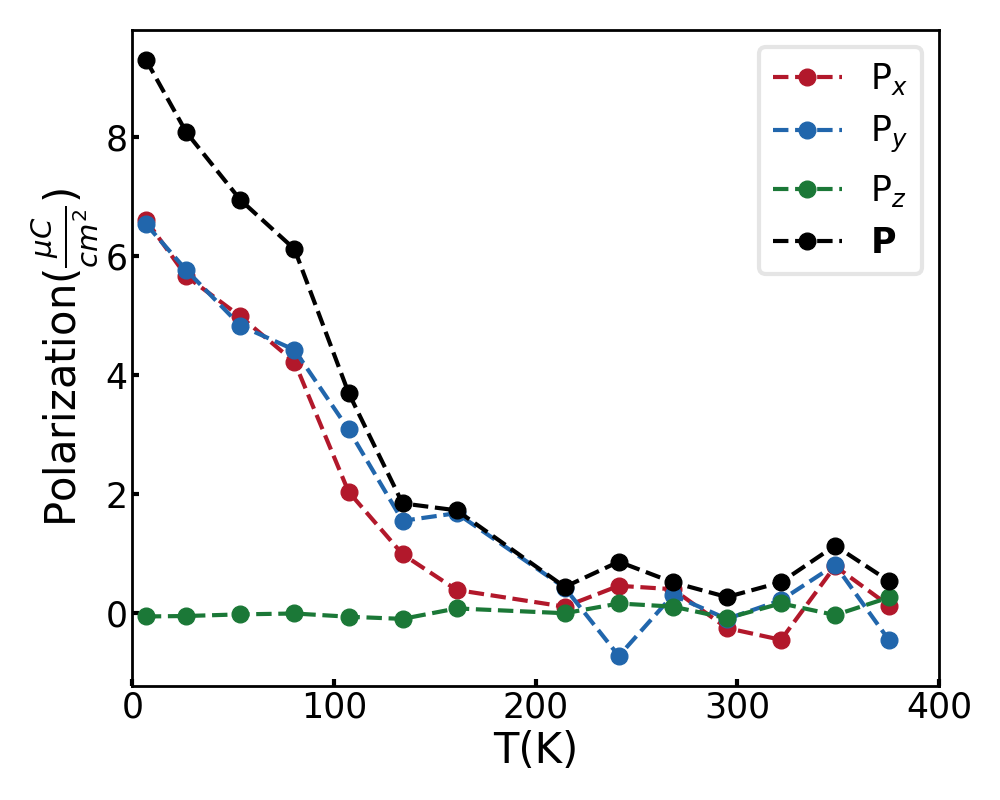}
\put(4.5,82){\Large\textbf{(a)}}
\end{overpic}
\begin{overpic}[width=0.47\linewidth]{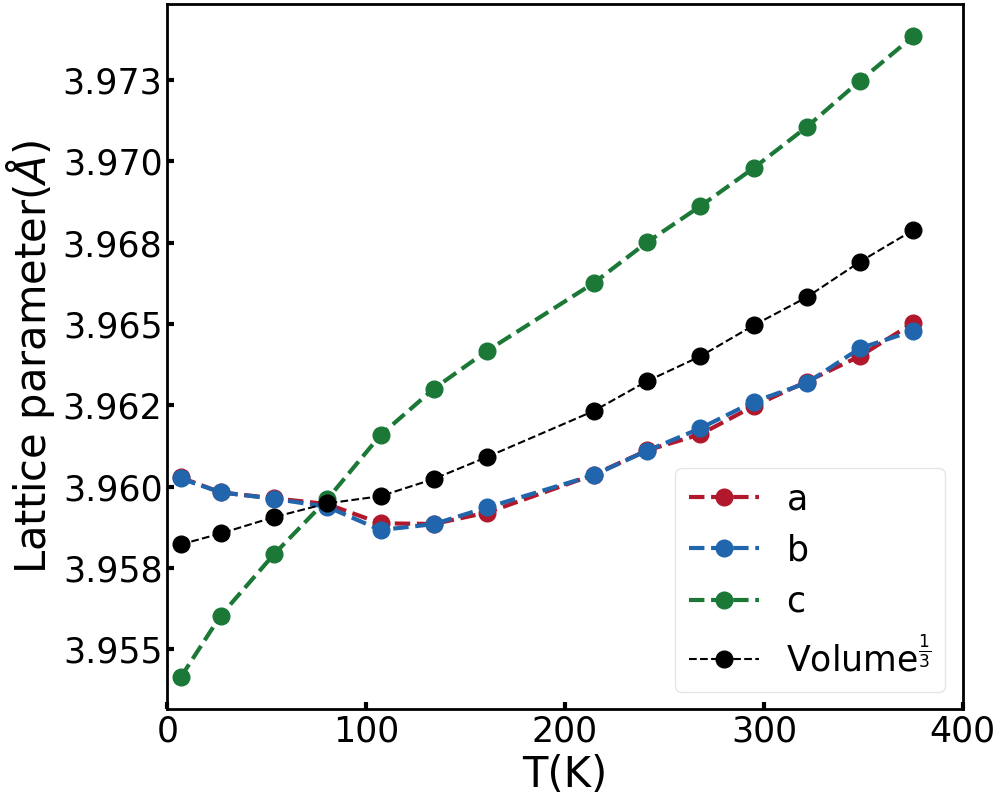}
\put(4.7,82){\Large\textbf{(b)}}
\end{overpic}

\caption{Temperature dependence of the average squared polarization (a) and the average lattice parameters (b) of a configuration with $N_z = 4$.}
\label{fig:pola}
\end{figure*}

The temperature evolution of selected properties for a representative $N_z=4$ configuration is shown in Figure~\ref{fig:pola}. A similar qualitative behavior is observed for all systems with $3 \le N_z < 6$. At low temperatures, the in-plane polarization dominates and persists up to approximately 135 K, where the polarization magnitude vanishes, as shown in Figure~\ref{fig:pola}a. Above this temperature, the system progressively enters a paraelectric-like regime, with the net polarization of all three components approaching zero and becoming increasingly governed by thermal fluctuations. This temperature also marks a crossover in the lattice parameters (Figure~\ref{fig:pola}b), where the unit cells evolve from an orthorhombic-like to a tetragonal-like distortion. All quantities were obtained by averaging the polarization and lattice parameters over the entire simulation cell at each temperature.

\vspace{0.1cm}

\section{Structural analysis}
To complement the topological analysis in the main text, a simple polarization-based criterion is used to assign the local structural character of each unit cell throughout the simulations. The thresholds to distinguish different polarization orientations are chosen to give a clear and physically reasonable separation. Although more sophisticated schemes exist \cite{Sepliarsky_2025}, this minimal approach is sufficient to capture the structural trends relevant here.

\begin{figure*}[ht]
\centering

\begin{overpic}[width=0.47\linewidth]{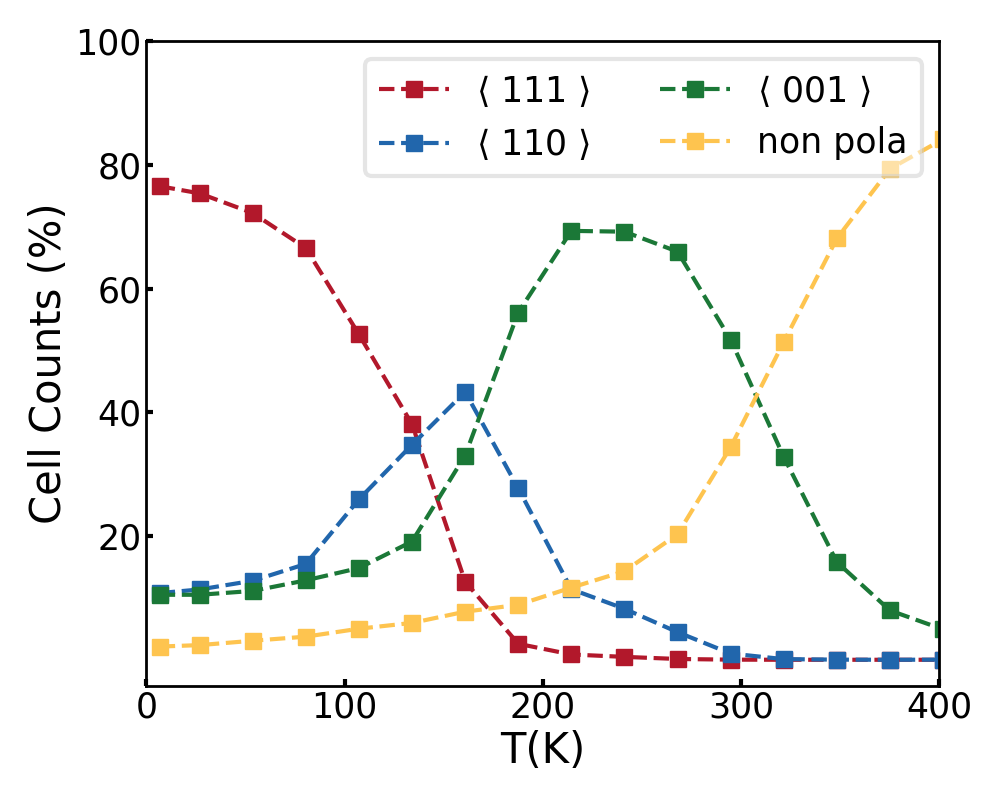}
\put(4.5,82){\Large\textbf{(a)}}
\end{overpic}
\begin{overpic}[width=0.47\linewidth]{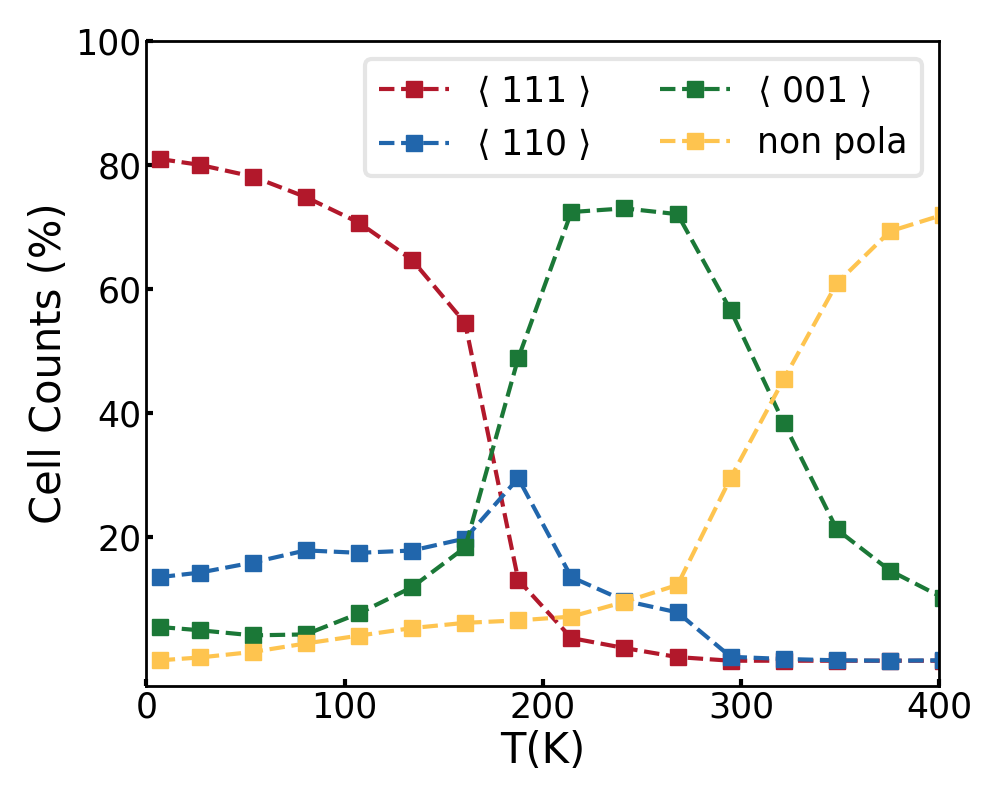}
\put(4.7,82){\Large\textbf{(b)}}
\end{overpic}

\caption{Exemplary $N_z=12$ system structural evolution. (a) Starting from a chiral bubble state, the system follows the bulk-like rhombohedral–orthorhombic–tetragonal–paraelectric sequence under confinement. (b) Starting from wave-helix state, the orthorhombic phase does not clearly emerge.}
\label{fig:phasetransition}
\end{figure*}

All simulated free-standing BTO layers with thickness $N_z \ge 6$ display the same qualitative behavior discussed in the manuscript. Figure~\ref{fig:phasetransition} shows the $N_z=12$ system as a representative example, illustrating how this polarization-based analysis complements the topological characterization of the emergent textures.

\vspace{0.35cm}

\section{Flux-closure domains: Vortex-Antivortex textures}
\begin{figure}[ht]
\centering
\includegraphics[width=0.56\linewidth]{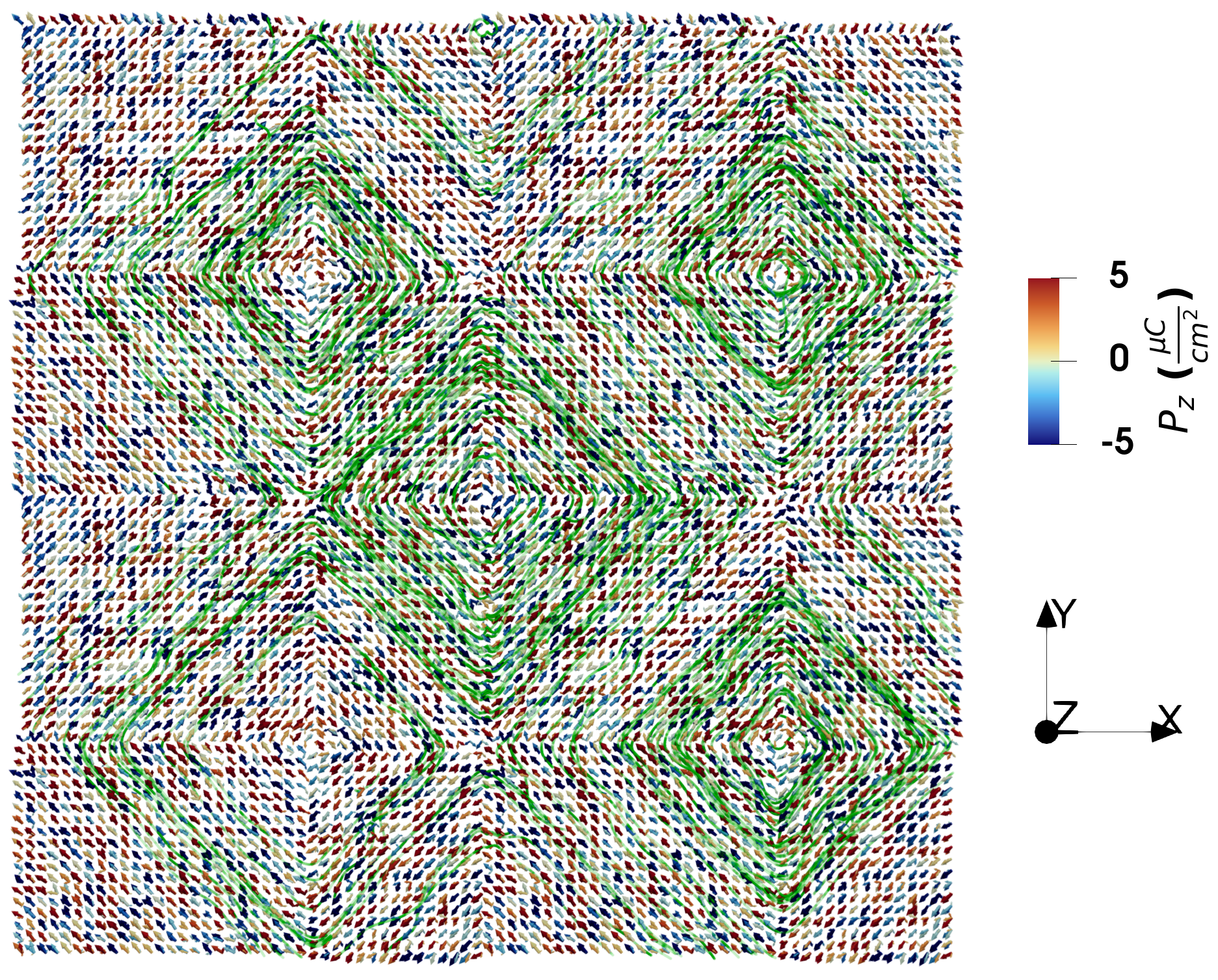}
\caption{\label{fig:flux} Top view of the polarization pattern in the ultrathin limit. The vector field displays a periodic array of vortices and antivortices. Streamlines serve as a guide to the eye to highlight the flux-closure nature of the domains, which lack an out-of-plane polarization component.}
\end{figure}

The field-induced reorganization in ultrathin layers ($3 \le N_z < 6 $) results in the formation of robust flux-closure patterns, as illustrated in Figure~\ref{fig:flux}. This top-view map of the local polarization reveals a periodic arrangement of vortices and antivortices.

While the in-plane projection of these textures may qualitatively resemble the cross-sections of the chiral bubble states discussed in the main text, their three-dimensional description is fundamentally different. In the chiral bubble phase (see Figure~2c of the manuscript), the polarization exhibits a strong out-of-plane component (P$_z$), leading to the formation of asymmetric domains with a well-defined vertical orientation. In contrast, the flux-closure states shown here are characterized by a polarization vector that remains strictly confined to the layer plane, with a negligible out-of-plane contribution. This distinction confirms the essentially two-dimensional nature of the vortex-antivortex network in the ultrathin regime.

\vspace{0.35cm}
\centering

\providecommand{\latin}[1]{#1}
\makeatletter
\providecommand{\doi}
  {\begingroup\let\do\@makeother\dospecials
  \catcode`\{=1 \catcode`\}=2 \doi@aux}
\providecommand{\doi@aux}[1]{\endgroup\texttt{#1}}
\makeatother
\providecommand*\mcitethebibliography{\thebibliography}
\csname @ifundefined\endcsname{endmcitethebibliography}  {\let\endmcitethebibliography\endthebibliography}{}